\documentclass[twocolumn,english,superscriptaddress]{revtex4-1}
\usepackage[T1]{fontenc}
\usepackage[latin9]{inputenc} 
\usepackage{tipa}
\usepackage{tipx}
\usepackage{graphicx}
\usepackage[dvipsnames]{xcolor}
\usepackage{float}

\makeatletter
\usepackage{babel}

\makeatother

\usepackage{babel}
\begin{document}

\title{Can a large packing be assembled from smaller ones?}
\author{Daniel Hexner}
\email{danielhe2@uchicago.edu}
\affiliation{The James Franck Institute and Department of Physics, The University
of Chicago, Chicago, IL 60637, USA}
\affiliation{Department of Physics and Astronomy, The University of Pennsylvania,
Philadelphia, PA, 19104, USA }
\author{Pierfrancesco Urbani}
\affiliation{Institut de Physique Th\'eorique, Universit\'e Paris
Saclay, CNRS, CEA, F-91191, Gif-sur-Yvette}
\author{Francesco Zamponi}
\affiliation{Laboratoire de Physique de l'Ecole Normale Sup\'erieure, CNRS, Paris,
France}
\affiliation{Universit\'e PSL, Sorbonne Universit\'e, Universit\'e Paris-Diderot,
Sorbonne Paris Cit\'e, Paris, France}

\begin{abstract}
We consider zero temperature packings of soft spheres, that undergo
a jamming to unjamming transition as a function of packing fraction.
We compare differences in the structure, as measured from the contact
statistics, of a finite subsystem of a large packing to a whole packing
with periodic boundaries of an equivalent size and pressure. We find
that the fluctuations of the ensemble of whole packings are smaller
than those of the ensemble of subsystems. Convergence of these two
quantities appears to occur at very large systems, which are usually
not attainable in numerical simulations. Finding differences between
packings in two dimensions and three dimensions, we also consider
four dimensions and mean-field models, and find that they show similar
system size dependence. Mean-field critical exponents appear to be
consistent with the 3d and 4d packings, suggesting they are above
the upper critical dimension. We also find that the convergence as
a function of system size to the thermodynamic limit is characterized
by two different length scales. We argue that this is the result of
the system being above the upper critical dimension. 
\end{abstract}
\maketitle
A starting point for characterizing the structure of ordered materials
are their microscopic subunits, or building blocks. Crystalline materials,
in their ground state, are defined by a single unit cell repeating
throughout the system. Quasicrystalline materials, while aperiodic,
still have a rather small number of building blocks. 
In the case of disordered materials,
each subsystem is different because of geometrical frustration, and the multiplicity of different
subsystems is huge~\cite{KurchanLevine2010}. Nonetheless, it is
interesting to ask, how different is a subsystem from the whole packing
it composes? This question addresses, in part, the effect of boundaries,
correlations in the structure, and multiplicity of ground states.

\begin{figure}
\includegraphics[scale=0.35]{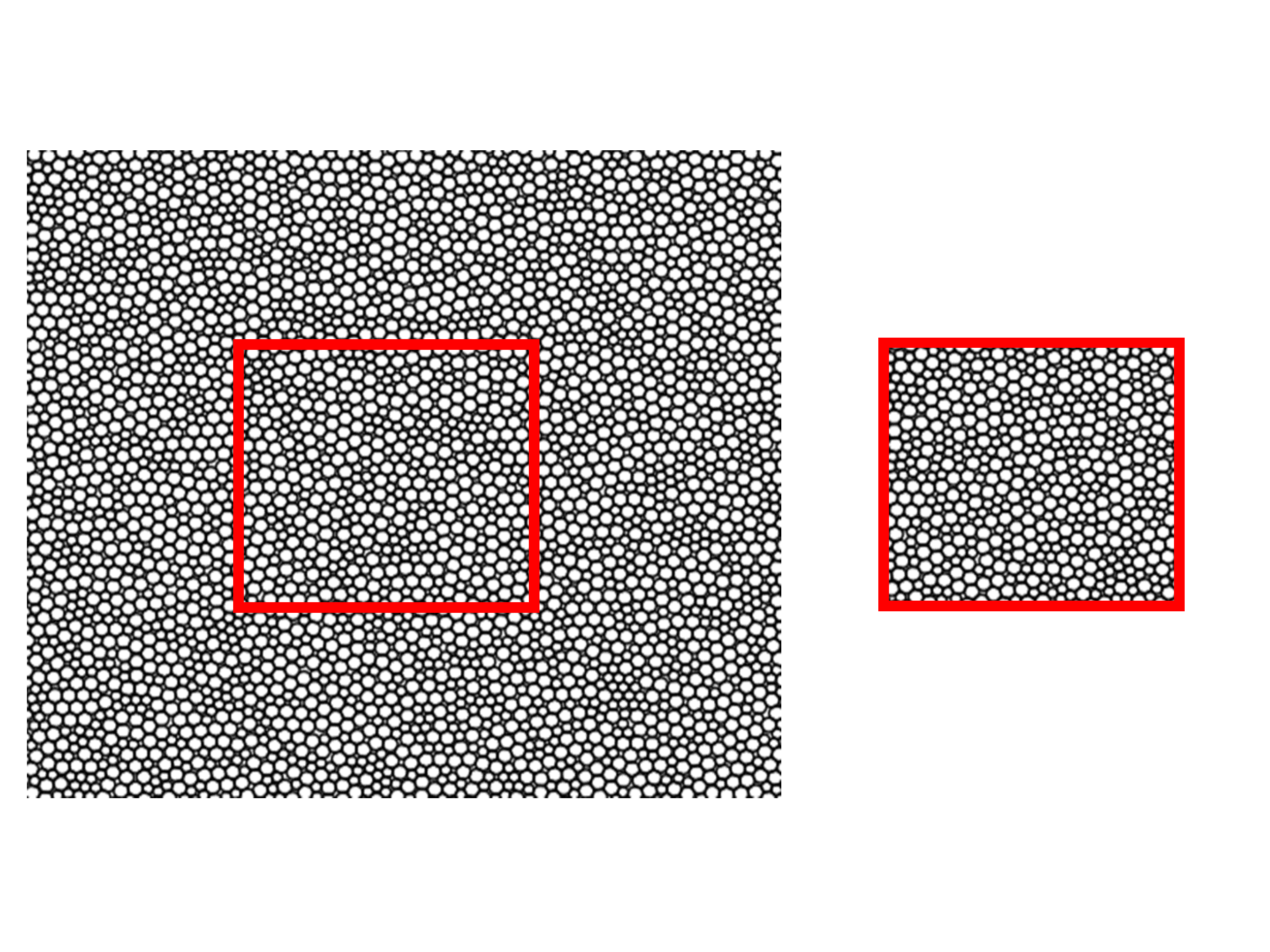} \caption{Illustration of the two ensembles: a subsystem of a large packing is
on the left, while a whole system of the same size is on the right.}
\label{fig:1} 
\end{figure}

In this paper, we ask this aforementioned question in a commonly studied
model for amorphous solids: disordered packings of soft spheres at
zero temperature~\cite{Wyart2005Ann,van2009jamming,liu2010jamming}.
This model undergos a rigidity transition, as a function of the packing
fraction~\cite{OHern2002}. We compare the ensemble of subsystems cut
out from large packings, to the ensemble of whole systems of the same
volume with periodic boundary conditions (Fig.~\ref{fig:1}). Recently
it has been found that the contact statistics possess unusual long
range correlations near the transition~\cite{hexner2018two}. We
therefore focus on contact fluctuations to compare the
two ensembles.

While we expect convergence of the two ensembles for large enough
systems, the system size $V^{*}$ for which these converge appears
to be in many cases well beyond that accessible via simulations.
For system sizes smaller than $V^{*}$, fluctuations in contacts
are significantly smaller in systems with periodic boundary conditions.
When approaching the jamming transition this disparity grows, and
$V^{*}$ appears to diverge, suggesting that it is associated with
a diverging length scale. To study convergence to the thermodynamic
limit we measure the contact fluctuations in whole systems as a function
of the distance to the jamming transition. We perform finite size
scaling to identify a length scale, and find that it differs from
those previously measured in the contact statistics. Finding differences
between 2d and 3d, we also consider 4d and mean-field variants of
the jamming model. These appear to be consistent with results from
3d, suggesting that the the upper critical dimension is below three~\cite{Goodrichfinitesize}.

To summarize, in this paper {\it (i)}~we introduce an entirely new procedure (in the context of jamming)
for finite size scaling analysis, by comparing subsystems of a large packing with periodic packings of the same size;
{\it (ii)}~we identify a new length scale, in addition to the ones reported in~\cite{hexner2018two};
{\it (iii)}~we provide accurate measurements of this new length scale
in dimension $d=2,3,4$ (note that previous calculations of jamming length scales could only obtain accurate
results for 2d systems);
{\it (iv)}~thanks to this improved finite size scaling analysis, we obtain strong quantitative
evidence that $d_u \sim 2$ is the upper critical dimension;
{\it (v)}~we analyze models with hypostatic jamming
and find that the corresponding suppression of fluctuations does not take place.
Our analysis thus shows
that anomalous contact fluctuations survive in mean-field like models,
and are crucially related to isostaticity.

We begin by defining the jamming model, in which overlapping particles
of radius $R_{i}$ interact via a harmonic potential: 
\begin{equation}
U_{ij}=\frac{1}{2}k\left(1-\frac{r_{ij}}{R_{i}+R_{j}}\right)^{2}\Theta\left(R_{i}+R_{j}-r_{ij}\right).\label{eq:jamming_potential}
\end{equation}
Here, $r_{ij}$ is the distance between the centers of the particles
and the Heaviside step function, $\Theta\left(x\right)$, insures
that only overlapping particles interact. In 2d the radii are chosen
to be polydisperse  \footnote{The radii are uniformly distributed in the range $[R_{min},R_{max}]$, with   $R_{max}/R_{min}=1.4$. }  to avoid crystallization, while in higher dimension
monodisperse particles lead to amorphous packings. 
We begin with particles distributed randomly and uniformly throughout space,
and minimize the energy at a constant pressure via FIRE algorithm~\cite{FIRE}
until the system reaches force balance.

An important quantity in understanding the geometry of the jamming
transition is the \emph{average} coordination number, $Z$. At the
marginally rigid state at zero pressure, the average coordination
number $Z$ attains a universal value that approaches $Z_{c}=2d$
for infinite system~\cite{alexander1998amorphous,edwards1998equations,moukarzel1998isostatic,Goodrichfinitesize}.
This amounts to the smallest number of contacts to maintain rigidity.
The excess coordination number $\Delta Z=Z-Z_{c}$, also characterizes
the distance from the jamming transition. Recently, it has been realized
that the coordination number possess subtle spatial correlations~\cite{hexner2018two,Wyart2005Ann}.
Unlike equilibrium critical systems that have diverging fluctuations
(associated with a diverging susceptibility), packings have anomalously
small fluctuations. At the jamming transition the bulk contact fluctuations
vanish, and the fluctuations inside a subsystem scale as its surface.

We now briefly review the metrics and the results of Ref.~\cite{hexner2018two}
for characterizing the fluctuations. 
For a packing of $N$ particles excluding the rattlers, we
define $Z_{i}$ to be the number
of particles in contact with particle $i$, $Z = \frac{1}{N} \sum_{i=1}^N Z_i$ the average
contact number in a given packing,
and $\delta Z_{i}=Z_{i}-Z$ the deviation from
the average. The fluctuations are then characterized
by measuring the variance in a hyper-cube of volume $\ell^{d}$, 
\begin{equation}
\sigma_{Z}^{2}\left(\ell\right)=\frac{1}{\ell^{d}}\overline{\big\langle \big(\sum_{i\in\ell^{d}}\delta Z_{i}\big)^{2}\big\rangle} .\label{eq:sigma_z_definition}
\end{equation}
Here the average $\langle \bullet \rangle$ is over many subsystems of a large packing, and the average $\overline{\bullet}$ is
over many large packings all at the same pressure, realized by different initial particle
positions prior to the energy minimization. If $\delta Z_{i}$ were
uncorrelated random variables, then $\sigma_{Z}^{2}\left(\ell\right)$
would not depend on $\ell$, since the number of particles scales
as the volume. At the jamming transition, $\Delta Z=0$, the fluctuations
scale in the smallest possible way, $\sigma_{Z}^{2}\left(\ell\right)\propto\ell^{-1}$,
implying that the sub-extensive fluctuations are dominated by the
surface of the enclosure~\cite{hexner2018two}. Such kind of fluctuations are called ``hyperuniform''~\cite{Torquato2003}
and have been observed in several strongly constrained physical systems ~\cite{Donev_2005,Berthier2015,Jancovici1981_onecomponent_plasma,HexnerHyper2015,tjhung2015hyperuniform,Bartolo2,Goldfriend2017,lei2019nonequilibrium,Torquato2018}.
At a finite distance
from jamming, $\Delta Z>0$, the fluctuations are only suppressed
up to a length scale $\xi_{f}\propto\Delta Z^{-\nu_{f}}$; above $\xi_{f}$,
the lack of correlations imply that $\sigma_{Z}^{2}\left(\ell\right)$
is independent of $\ell$. We also remark that unlike typical critical
systems, here there are two different diverging length scales. A second
length scale, $\xi_{Z}\propto\Delta Z^{-\nu_{z}}$, can be measured
from $\overline{\left\langle \delta Z\left(r\right)\delta Z\left(0\right)\right\rangle} $
and is different than $\xi_{f}$, having different exponents $\nu_{f}>\nu_{z}$~\cite{hexner2018two}.

In this paper, in addition to $\sigma_{Z}^{2}\left(\ell\right)$,
we characterize sample-to-sample fluctuations of many jamming configurations
with periodic boundaries, at the same value of pressure. We define
\begin{equation}
\delta^{2}Z\left(N\right)\equiv N\left[\overline{ Z^{2} } -\overline{ Z}^{2}\right],\label{eq:dZ_sample_definition}
\end{equation}
where the average $\overline{\bullet}$ is over distinct packings at constant pressure, 
and $Z=\frac{1}{N}\sum_{i=1}^{N}Z_{i}$, as before.
The factor of $N$ on the right hand side of Eq.~(\ref{eq:dZ_sample_definition})
ensures convergence to a finite value in the limit of $N\rightarrow\infty$.
We also note that in the infinite size limit, where boundary condition
are unimportant, $\rho\delta^{2}Z(N=\infty)=\sigma_{Z}^{2}\left(\ell=\infty\right)$,
where $\rho$ is the number density. To see how this is related to
$\xi_{f}$, we note that a sufficiently large system can decomposed
into uncorrelated sub-regions of volume $\xi_{f}^{d}$. In each such
region the fluctuations scale as the surface, $\xi_{f}^{d-1}$. Because
the number of uncorrelated regions is $\left(L/\xi_{f}\right)^{d}$,
we obtain 
\begin{equation}
\delta^{2}Z\left(N\gg\xi_{f}^{d}\right)\propto\frac{1}{\rho}\xi_{f}^{-1}\propto\Delta Z^{\nu_{f}}.\label{eq:inf_limit}
\end{equation}

\begin{figure}[t]
\includegraphics[scale=0.55]{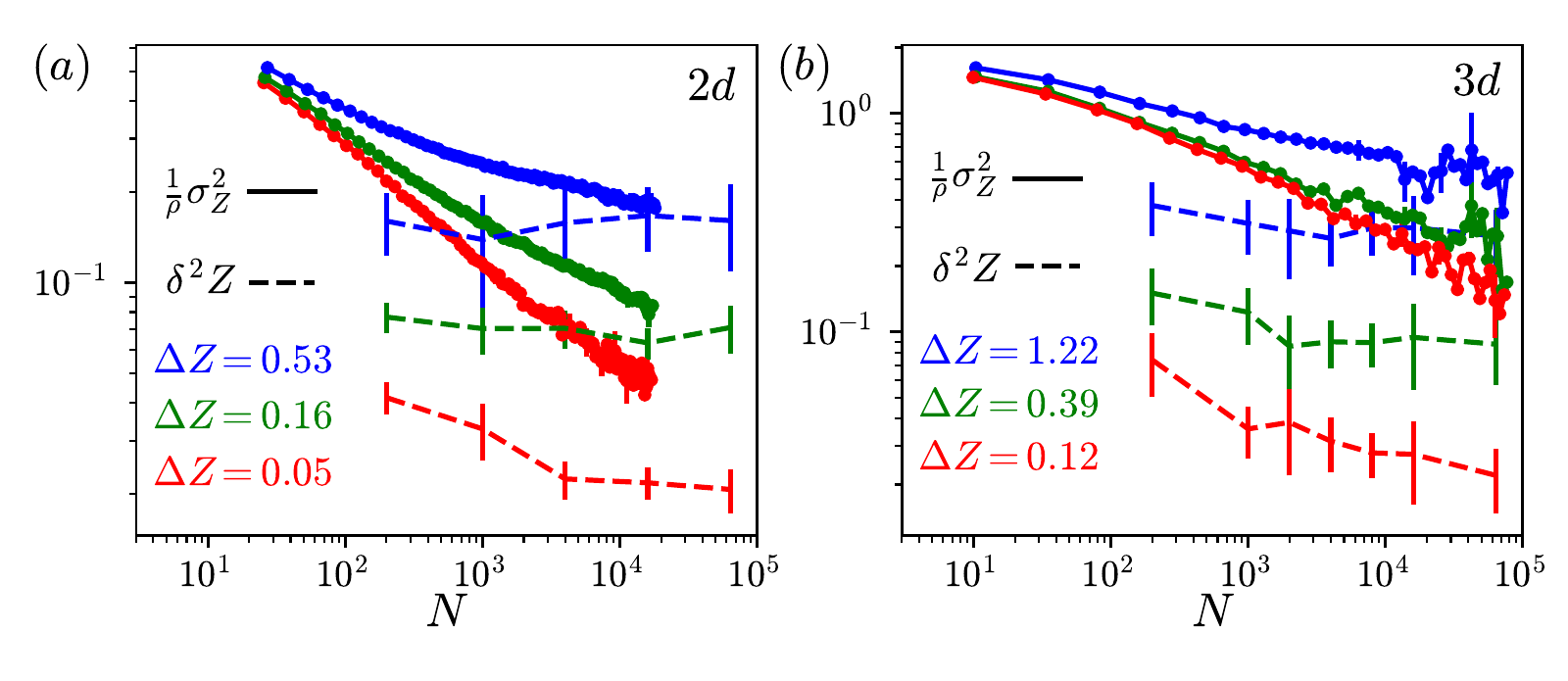} 
\caption{Comparison of the fluctuations of contacts of entire packings (full line), $\delta^{2}Z$,
to the contact fluctuations of a sub-system (dashed line) $\sigma_{Z}^{2}/\rho$
with the same number of particles in 2d (a) and 3d (b).
For $\sigma_{Z}^{2}/\rho$, error bars are shown only for some points to ease visualization.
}
\label{fig:sig_vs_dZ} 
\end{figure}

We now turn to show how $\sigma_{Z}^{2}\left(\ell\right)$ and $\delta^{2}Z(N)$
compare on a finite length scale. To make this comparison, we plot
$\sigma_{Z}^{2}\left(\ell\right)/\rho$ as a function of $N=\rho\ell^{d}$. 
Results in 2d and in 3d (Fig.~\ref{fig:sig_vs_dZ})
show that for finite $N$ sample-to-sample fluctuations, $\delta^{2}Z$, are smaller
than $\sigma_{Z}^{2}/\rho$ and have a fairly weak dependence on system
size. Because
$\sigma_{Z}^{2}\left(\ell\right)/\rho$ and $\delta^{2}Z(N)$ converge
in the thermodynamic limit, one can define a length scale at which
the two ensembles converge. 
This length scale is surprisingly large,
especially in 3d,
and the difference between the two ensembles grows dramatically
upon approaching the jamming transition, $\Delta Z\to0$. 
Even at $\Delta Z\approx1.22$, which is usually
considered to be far from the jamming transition, convergence occurs
for $N>10^{5}$. For $\Delta Z\approx0.12$ convergence can be extrapolated
to occur for system sizes of $N>10^{8}$ which are currently not attainable
numerically. The fact that convergence in 2d appears to occur at smaller
$N$ suggests that it is dominated by a correlation length.

It is interesting to speculate on why the sample-to-sample fluctuations
are much smaller than the fluctuations in a subsystem. Subsystems,
by definition, have boundaries, and these always entail a surface
contribution that scales as $\ell^{-1}$. This suggests that $\delta^{2}Z$,
which lacks these surface fluctuations, measures the bulk contribution
to the fluctuations, scaling as $\Delta Z^{\nu_{f}}$, as noted in
Eq.~(\ref{eq:inf_limit}). We will demonstrate that this scaling
holds for both ensembles on length scales larger than $\xi_{d}=\Delta Z^{-\nu_{d}}$,
and below this length scale finite size effects are present. The length
$\xi_{d}$ is smaller than both $\xi_{Z}$ and $\xi_{f}$ in $d>2$,
so that $\nu_{d}<\nu_{z}<\nu_{f}$.

\begin{figure}
\includegraphics[scale=0.55]{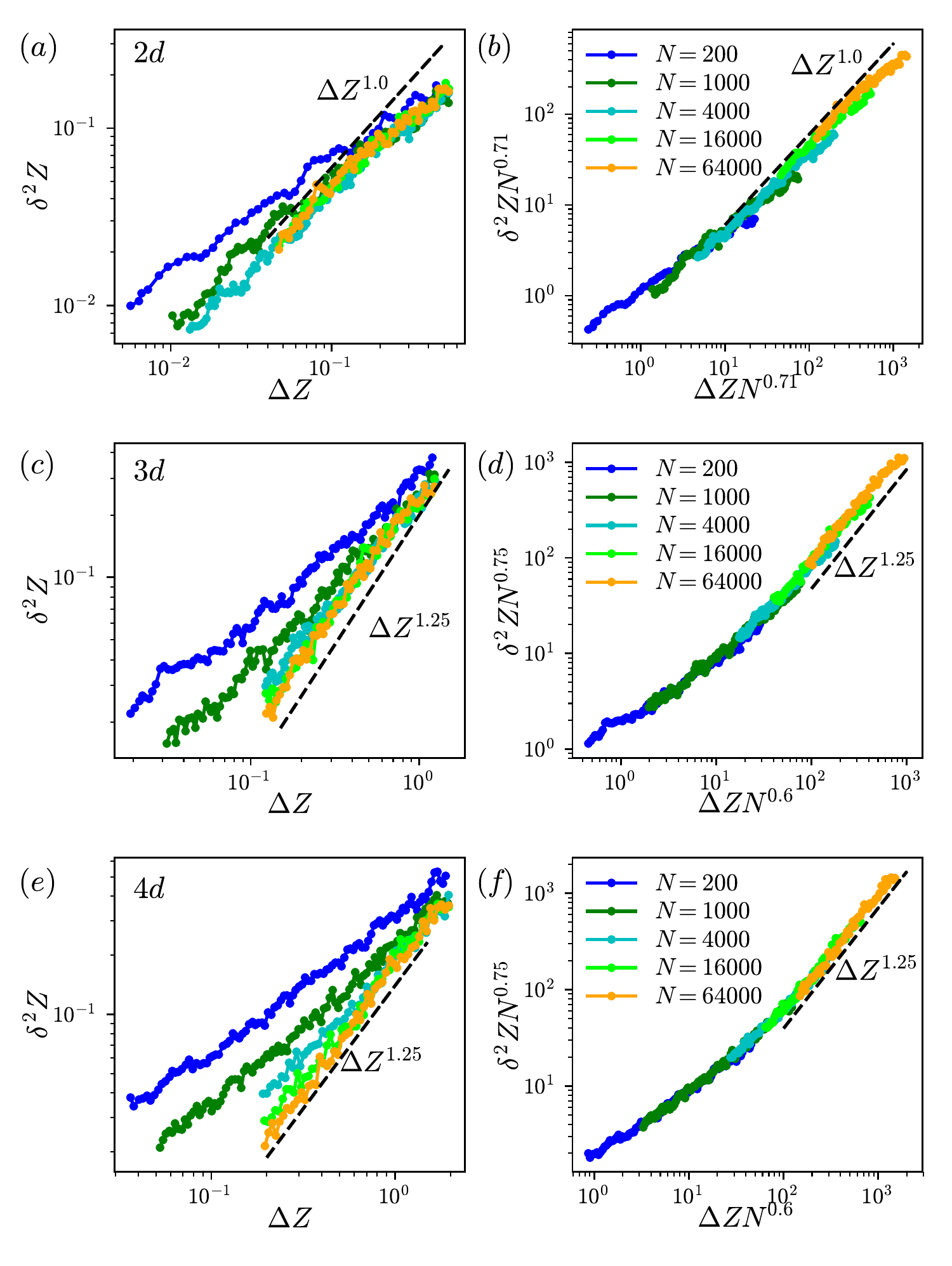} \caption{$\delta^{2}Z$ as a function of $\Delta Z$ in two (a), three (c)
and four dimensions (e). The curves are collapsed by rescaling the
axis with different power of $N$ in two (b), three (d) and four dimensions
(f). }
\label{data_2d_and_3d} 
\end{figure}

In order to measure $\nu_{d}$ and $\nu_{f}$, in Fig.~\ref{data_2d_and_3d}
we plot $\delta^{2}Z$ as a function of $\Delta Z$,
for different system sizes, in $d=2,3,4$. In all cases there is a
dependence on system size, mostly observed at small values of $\Delta Z$,
where $\delta^{2}Z$ decreases with system size. This effect seems
to be smaller in 2d than in 3d and 4d. For large enough systems the
curves appear to converge, and for these we measure $\nu_{f}$, from
Eq.~(\ref{eq:inf_limit}). The values of $\nu_{f}$ are consistent
with Ref.~\cite{hexner2018two}, where $\nu_{f}^{2d}\approx1.0$ and $\nu_{f}^{3d}\approx1.25$  (see Appendix for a comparison between 2d and 3d). 
The variation between 2d and 3d is
also manifest in the qualitative shape of the curves: unlike the 3d
curves, the 2d results taper off at large values of $\Delta Z$. The
behavior in 4d is consistent with the 3d case, $\nu_{f}^{4d}\approx1.25$.
To characterize the system size dependence, we collapse the different
curves by assuming a scaling form, 
\begin{equation}
\delta^{2}Z=N^{-\beta}f\left(\Delta ZN^{\alpha}\right).\label{eq:scaling_form}
\end{equation}
Requiring that in the limit of $N\rightarrow\infty$, $\delta^{2}Z\propto\Delta Z^{\nu_{f}}$
is independent of the system size, yields $\beta=\alpha\nu_{f}$.
Hence, given $\nu_{f}$, the curves can be collapsed by varying a
single exponent. Because the number of particles is proportional to
the volume, $L^{d}$, the argument of Eq.~(\ref{eq:scaling_form})
can be rewritten as $\Delta ZN^{\alpha}\propto\left(\frac{L}{\xi_{d}}\right)^{\alpha d}$,
with $\xi_{d}\equiv\Delta Z^{-\nu_{d}}$ and $\nu_{d}\equiv1/(\alpha d)$.
The collapse shown in Fig.\ref{data_2d_and_3d} yields approximately
$\alpha\approx0.6$ both in 3d and 4d, implying that $\nu_{d}$ depends
on dimension. In 3d the exponent $\nu_{d}$ is smaller than $\nu_{z}$
and $\nu_{f}$, measured in Ref.~\cite{hexner2018two}. In 2d it
is difficult to determine $\alpha$; because $\nu_{f}^{2d}\approx1.0$
there is a range of exponents $\alpha$$=$$\beta$ which collapse
the data reasonably well.

\begin{figure}
\includegraphics[scale=0.55]{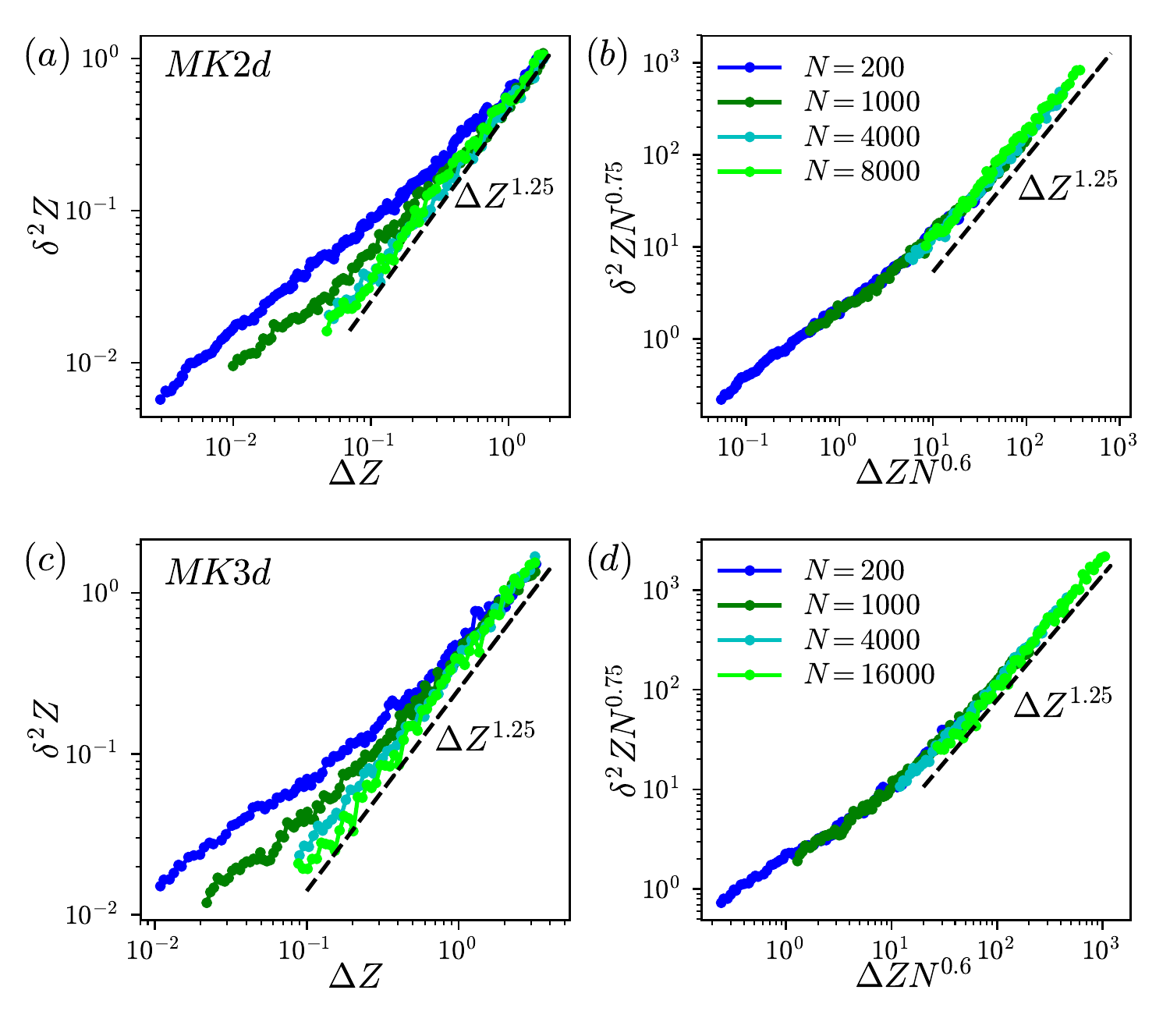} \caption{$\delta^{2}Z$ as a function of $\Delta Z$ for the MK mean-field
model in (a) 2d and (c) 3d. The curves are collapsed by rescaling
the axis with different powers of $N$ in (b) and (d), respectively.}
\label{mean_field_models} 
\end{figure}

We also consider the mean-field
limit of the jamming model, by simulating the Mari-Kurchan (MK) model~\cite{mari2011dynamical}.
This model retains most of the details of the jamming model, including
the interaction potential in Eq.~(\ref{eq:jamming_potential}), but
aims at disrupting the spatial correlations by varying the spatial
metric. A given particle $i$ sees particle $j$ at a location shifted
by a random value, $d_{ij}$, which is uniformly distributed through
space. The potential between particle $i$ and $j$ is thus given
by $U_{ij}=U\left(\left|r_{i}-r_{j}-d_{ij}\right|\right)$, and the
interaction between particles does not depend on the actual Euclidean
distance between them. As a consequence, spatial dimension $d$ is not
expected to play any role in the criticality, and the model is mean-field.

Fig.~\ref{mean_field_models} shows the results of the simulations
of the MK model in 2d and 3d, which overall appear very similar to
the 3d and 4d non-mean field variant. In the large system size limit,
the MK curves appear to converge to $\Delta Z^{\nu_{f}^{MK}}$ and
$\nu_{f}^{MK}\approx1.25$ independently of dimension. The exponent
is also very similar to the 3d and 4d result, suggesting that the
upper critical dimension is less than three~\cite{Goodrichfinitesize}.
The MK curves can be collapsed using
the scaling form in Eq.~(\ref{eq:scaling_form}), with exponents agreeing within errors 
with those of the 3d and 4d jamming model.
Hence, the collapse of $\delta^{2}Z$ does not depend on the length
of the system but rather its volume, or number of particles. Because
$\alpha$ is independent of dimension (for 3d and above), the scaling
of the length scale $\xi_{d}\propto\Delta Z^{-\nu_{d}}$, with $\nu_{d}=1/\alpha d$,
does depends on dimension. This is contrasted by the collapse of the
fluctuations in a subsystem with $\ell/\xi_{f}$, the length $\xi_{f}$
scaling independently of dimension.

We now discuss the finite size scaling of $\delta^{2}Z$. The theory
of finite size scaling above the upper critical dimension $d_{u}$
that has emerged from work on the Ising model \cite{binder1985finite,wittmann2014finite}
predicts two types of scalings, depending on the quantities considered.
Finite size scaling of fluctuations of whole systems collapse as a
function of the system size (number of particles). The intuitive reason
is that some mean-field models cannot be embedded in Euclidean space,
and are thus defined by system size alone. Nonetheless, 
the scaling of quantities associated to a finite wavelength 
depend on the ratio of a length and the correlation length,
whose exponent is given by its mean-field value.
Hence, quite generally one can define two diverging length scales
for systems above $d_{u}$, in contrast to dimension below $d_{u}$
where only one correlation scale is relevant. In the jamming model
this scenario is realized by a different collapse of $\sigma_{Z}^{2}$
and $\delta^{2}Z$. One can exploit these different collapses to measure
$d_{u}$, by estimating the dimension where the two length scales
coincide. Extrapolating $\nu_{d}=1/(\alpha d)$ to the dimension where
it is equal to the mean field exponent yields $d_{u}$. However, as
mentioned above, in the jamming transition there are two length scales
that characterize contact statistics~\cite{hexner2018two}: $\xi_{Z}$,
which characterizes the two point correlation function, and $\xi_{f}$,
which characterizes the cross-over of the hyperuniform fluctuations
to the normal fluctuations. We argue that the first is a more fundamental
quantity, and using the exponent $\nu_{z}^{3d}\approx0.85$ measured
in~\cite{hexner2018two} we obtain that $\nu_{d}=\nu_{z}$ when $d_{u}\approx2$.
This estimate is consistent with our findings
that 2d appears somewhat different. In the latter case, $\delta^{2}Z$ would
collapse with the 2d correlation length, implying that
$\alpha_{2d}=1/(2\nu_{z}^{2d})$. The collapse in Fig.~\ref{data_2d_and_3d}
shows reasonable agreement, but the finite range of our data allows
other values as well. Note that it has long been thought that $d_u=2$~\cite{liu2010jamming,wyart2005rigidity}.
This is mostly based on observations that exponents appear to be independent
of dimension, for $d\geq2$~\cite{ohern2003jamming,charbonneau2015jamming,charbonneau2014fractal,charbonneau2017glass}.

We also consider a different class of jamming models, which are not
isostatic at the jamming transition~\cite{Brito2018PRX}, as it is the case for
packings of non-spherical particles~\cite{brito2018universality,Donev2007,donev2004improving}.
The inter-particle interactions are given by Eq.~(\ref{eq:jamming_potential}),
but the particle radii are also considered as degrees of freedom of ``breathing particles''.
To insure that particles do not shrink to zero, a confining potential
is assigned to the radii: 
\begin{equation}
U_{r}=\sum_{i}\frac{k_{r}}{2}\left(R_{i}-R_{i}^{0}\right)^{2}.
\end{equation}
To avoid crystallization, in 2d we consider a bidisperse particle
distribution, where the diameter $R_{i}^{0}$ of half of the particles
is larger by a factor $1.4$ than that of the others. An important
feature of this model is the scaling of the stiffness $k_{r}$. To
achieve a radii distribution with finite width at jamming, $k_{r}=Pk_{0}$,
where $P$ is the pressure and $k_{0}$ sets the overall magnitude
of the stiffness. In the limit of $k_{0}\rightarrow\infty$, the behavior
of the usual jamming transition is recovered.

\begin{figure}
\includegraphics[scale=0.45]{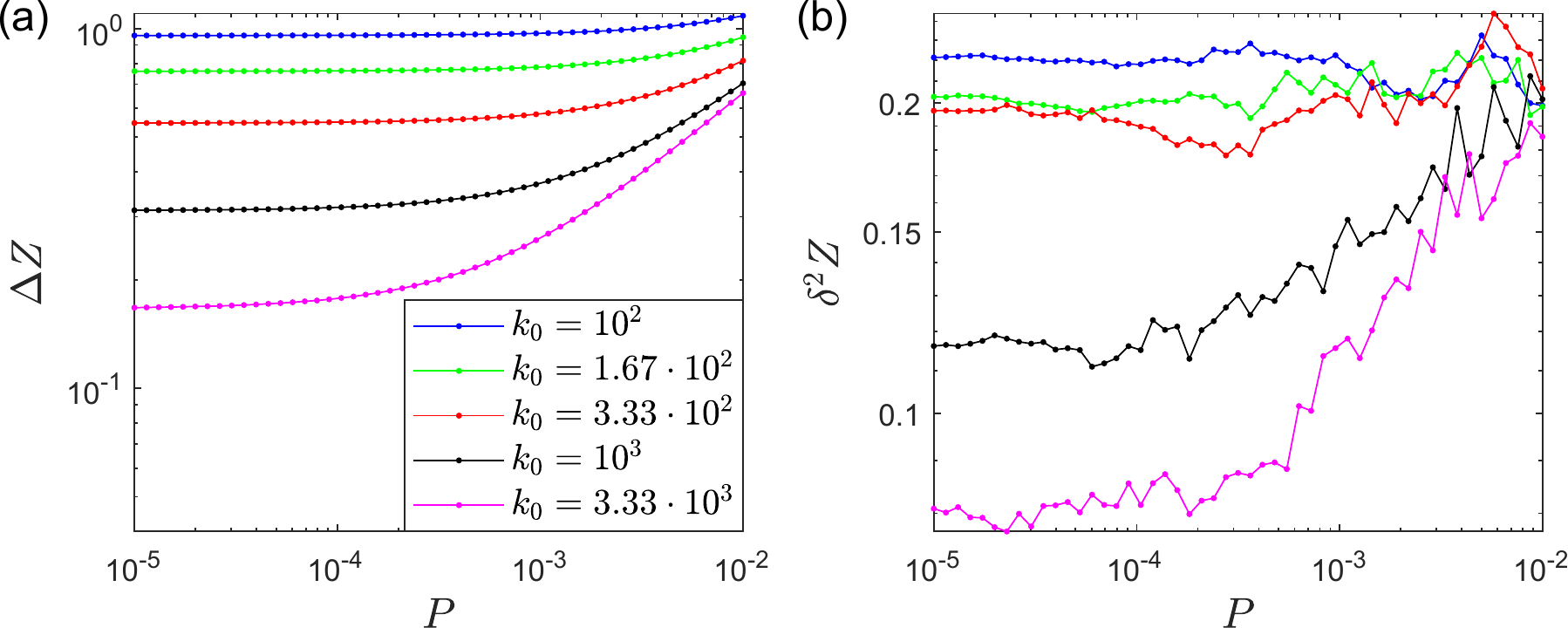} 
\caption{The average $\Delta Z$ (a) and fluctuations $\delta^{2}Z$ (b) as
a function of pressure for the breathing particle model. Note that unlike
the usual jamming transition, the contact fluctuations are non-zero
at the transition. The system is two dimensional with $N=4000$ particles. }
\label{fig:ZvsP} 
\end{figure}

Unlike the usual jamming transition, in the limit of $P\rightarrow0$,
the system is not isostatic, $\Delta Z>0$. In Fig.~\ref{fig:ZvsP}(a)
we show that our simulations are consistent with the results of Ref.~\cite{Brito2018PRX}.
In Fig.~\ref{fig:ZvsP}(b) we plot the contact fluctuations as a
function of pressure. In the limit of $P\rightarrow0$, the contact
fluctuations $\delta^{2}Z$ remain finite, in contrast to the vanishing
fluctuations of the conventional jamming transition. This suggests
that isostaticity is crucial for the suppression of fluctuations
and the divergence of correlation lengths.

We conclude by reiterating our main results. Fluctuations in whole periodic
systems are smaller than in subsystems of the same size. This is most significant at
the jamming transition where $\delta^{2}Z\rightarrow0$, while
the fluctuations in a subsystem scale as $\ell^{-1}$, which is the
fastest possible decay. Moreover, $\delta^{2}Z$ and $\sigma_{Z}^{2}$
converge to their thermodynamic value with different exponents for
3d and above. The sample-to-sample fluctuations, $\delta^{2}Z$, approach
their asymptotic value at system sizes $N\sim (\xi_d)^d \sim \Delta Z^{-1/\alpha}$,
where $\alpha\approx0.6$ is independent of dimension. Fluctuations
in a subsystem only reach their asymptotic value at a system length
$\ell\sim\xi_f\sim \Delta Z^{-\nu_{f}}$, where $\nu_{f}\approx1.25$ is also
independent of dimension. Under typical system sizes and values of
$\Delta Z$, usually considered in simulations, the system is well
below $\xi_{f}$. It is therefore interesting to explore if new behaviors
arise for large systems and how $\xi_{f}$ and $\xi_{Z}$ affect the
behavior of the packings. As a byproduct of this analysis, we obtain
an estimate of the upper critical dimension $d_{u}\approx2$, as the
dimension where the distinction between $\xi_{d}$ and $\xi_{Z}$
disappears. Our results also show that a signature of suppressed fluctuations
survive in mean-field variants of the jamming model. The exact solution
of these models~\cite{charbonneau2014exact} enables, in principle,
the analytic calculation of $\nu_{f}$, although the calculation is technically
involved. The exponent $\nu_{z}$ that characterizes the spatial decay
of $\left\langle \delta Z\left(r\right)\delta Z\left(0\right)\right\rangle $
still remain inaccessible to present theory.

\textit{Acknowledgments --} We warmly thank Andrea Liu and Sid Nagel
for important discussions. This work was supported by a grant from
the Simons Foundation (\#348125, Sid Nagel and Daniel Hexner, and
\#454955, Francesco Zamponi) and from "Investissements d'Avenir"
LabEx PALM (ANR-10-LABX-0039-PALM).

\section*{Appendix}
\subsection*{Minimization at constant pressure}

In this section we present further details of how packings are prepared
at a constant pressure. As discussed in the main text the energy of
the system depends on the inter-particle distance: 
\begin{equation}
U_{ij}=\frac{1}{2}k\left(1-\frac{r_{ij}}{R_{i}+R_{j}}\right)^{2}\Theta\left(R_{i}+R_{j}-r_{ij}\right).
\end{equation}
Working at a finite pressure provides a tighter control over the distance
from the jamming transition, than at constant packing fraction. In
the latter, near the jamming transitions some of the packings may
be under constrained and some packings could be over constrained.
To maintain a constant pressure we minimize the enthalpy 
\begin{equation}
H=U+P_{0}V.
\end{equation}
Here the target pressure is $P_{0}$. We employ the FIRE minimization
algorithm, which evolves based in the gradients of energy (or enthalpy)
\cite{FIRE}. The volume of the box is also a coordinate that varies
during the minimization and its dynamics depend on the gradient with
respect to the  volume, $P=-\frac{\partial U}{\partial V}$. When H is a minimum  $\frac{\partial H}{\partial V}=0$, implying that $P=-\frac{\partial U}{\partial V}=P_0$. The
pressure of the system, $P$, is given by the diagonal of the virial
stress tensor:
\begin{equation}
\tau_{ij}=\frac{1}{V}\sum_{b}r_{b,i}f_{b,j}.
\end{equation}
 The sum is over all bonds, $V$ is the volume, $r_{b}$ is a vector
that connects the center of two interacting particles and $f_{b}$
is inter-particle force along and it is pointed in the same direction
as $f_{b}$. 

\subsection*{Definition of $\sigma_Z^2(\ell)$.}

The definition of $\sigma_Z^2(\ell)$ given in the main text is
\begin{equation}
\sigma_{Z}^{2}\left(\ell\right)=\frac{1}{\ell^{d}}\overline{\big\langle \big(\sum_{i\in\ell^{d}}\delta Z_{i}\big)^{2}\big\rangle} \label{sss}
\end{equation}
where
\begin{equation}
\delta Z_i = Z_i - Z
\label{ZZZ_i}
\end{equation}
and 
\begin{equation}
Z= \frac 1N \sum_{i=1}^N Z_i
\end{equation}
A small variant of Eq.~(\ref{sss}) is to replace $Z$ with $\overline{Z}$ which is the sample-to-sample average of $Z$. 
We denote the corresponding $\sigma_Z^2$ as $\overline \sigma_Z^2(\ell)$.
For $\ell\to \infty$ we have that 
\begin{equation}
\overline \sigma_Z^2(\ell\to \infty)/\rho \equiv \delta^2Z(N\to \infty)
\end{equation}
However, since for $N\to \infty$, $Z\to \overline{Z}>0$ we get that the difference between $\sigma_Z^2$ and $\overline \sigma_Z^2$ is a subleading term that vanishes for $N\to \infty$ and $\ell\to \infty$, so that
\begin{equation}
\sigma_Z^2(\ell\to \infty)/\rho \equiv \delta^2Z(N\to \infty)
\end{equation}
Finally the same argument holds if we replace 
$Z$ by its local average $Z_{(i)}$, meaning its average inside the box in which $Z_i$ is computed in Eq.~(\ref{ZZZ_i}).
Since for $\ell\to \infty$, $Z_{(i)}\to Z>0$ up to subleading corrections, one can interchange the definitions without affecting the large $\ell$ behavior.

\subsection*{Comparison of  $\nu_{f}$ in 2d and 3d}

In the main text we showed that in the large system limit $\delta^{2}Z\propto\Delta Z^{\nu_{f}}$,
where our data suggested that $\nu_{f}^{2d}\approx1.0$, while in
higher dimension $\nu_{f}\approx1.25$. To visualize these two possible
scalings we plot these two power-laws. We note that the exponents
are deduced based on the collapse of $\delta^{2}Z$, as well as the
collapse of $\sigma_{f}^{2}$ in Ref. \cite{hexner2018two}.

\begin{figure}[H]
\includegraphics[scale=0.6]{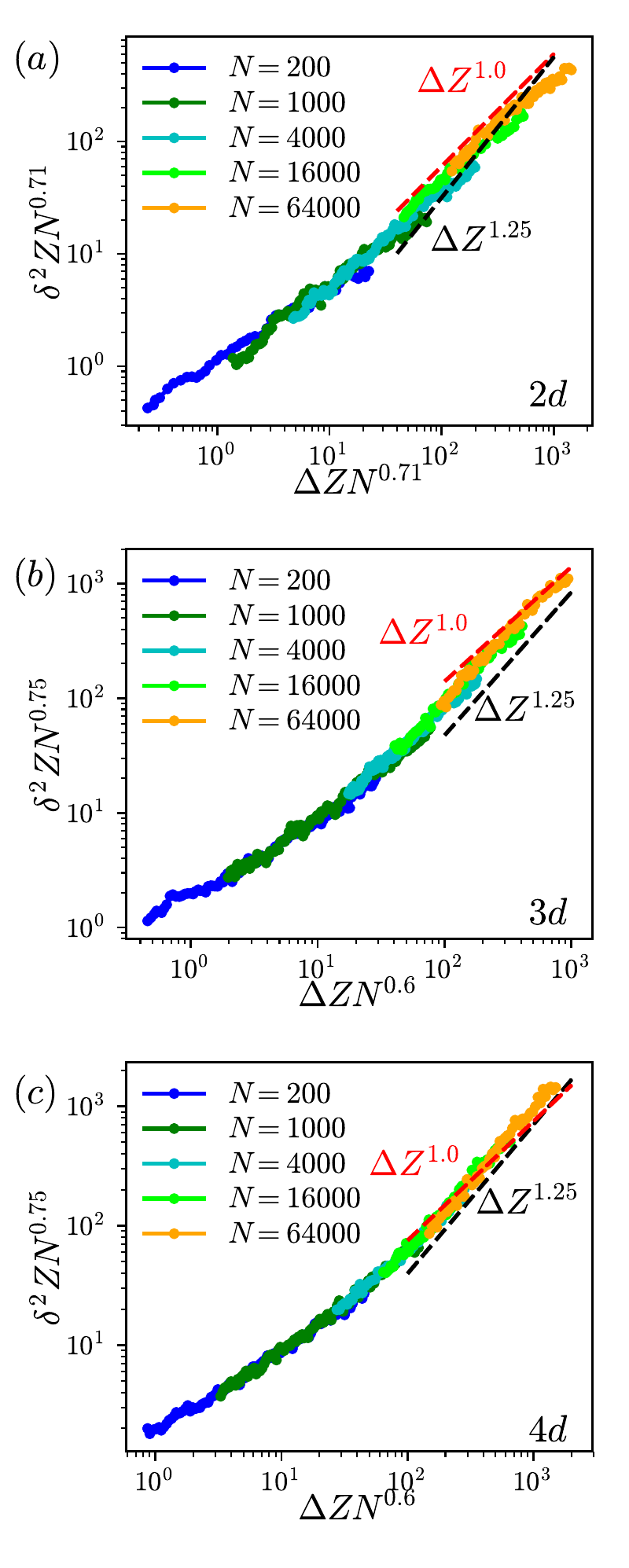}

\caption{A comparison of the two slopes, $\Delta Z^{1.0}$ and $\Delta Z^{1.25}$.
In two dimension the collapse suggests $\nu_{f}^{2d}\approx1.0$,
while in three and four dimensions $\nu_{f}\approx1.25$. }

\end{figure}

 \bibliographystyle{apsrev4-1}
\bibliography{biblo}

\end{document}